\documentclass{aa}  
\usepackage{float}
\usepackage{graphicx}
\usepackage{comment}
\usepackage{fontawesome}
\usepackage[dvipsnames]{xcolor}
\usepackage{txfonts}

\usepackage[pdfencoding=auto,psdextra]{hyperref}
\hypersetup{
    colorlinks=true,
    linkcolor=blue,
    filecolor=magenta,      
    urlcolor=blue,
    citecolor=blue
}
\usepackage{cleveref}
\crefname{chapter}{Chap.}{Chaps.}
\crefname{section}{Sect.}{Sects.}
\crefname{figure}{Fig.}{Figs.}
\Crefname{chapter}{Chapter}{Chapters}
\Crefname{section}{Section}{Sections}
\Crefname{figure}{Figure}{Figures}
\crefname{equation}{Eq.}{Eqs.}

\title{Separate Universe Super-Resolution Emulator}

\begin{document} 
   \author{Dennis Fremstad$^1$
        \and 
        Julian Adamek$^{2,3,4}$
        \and
        David F.\ Mota$^1$
    }
   \institute{$^1$Institute of Theoretical Astrophysics, University of Oslo,
PO Box 1029, Blindern 0315, Oslo, Norway \\
    $^2$Institut f\"ur Astrophysik, Universit\"at Z\"urich, Winterthurerstrasse 190, 8057 Z\"urich, Switzerland\\
    $^3$D\'epartement de Physique Th\'eorique, Universit\'e de Gen\`eve, 24 Quai Ernest-Ansermet, 1211 Gen\`eve 4, Switzerland\\
    $^4$Institut f\"ur Teilchen- und Astrophysik, ETH Z\"urich, Wolfgang-Pauli-Strasse 27, 8093 Z\"urich, Switzerland
    }
 
  \abstract
  {
  We present a machine-learning model for generating super-resolution $N$-body simulations with non-vanishing spatial curvature, conditioned on a given low-resolution field, $\Omega_k$, $\Omega_\mathrm{m}$, $\sigma_8$, $h$, and redshift. 
  By upscaling the resolution of $N$-body simulations, such models can drastically reduce the computational cost of producing high-resolution simulations suitable for modelling current and future surveys of large-scale structure. Our 
  model is trained as a generative adversarial network, allowing injected noise to be interpreted as stochastic structure and enabling the generation of an ensemble of plausible high-resolution realisations. 
  We evaluate the model performance by comparing key cosmological summary statistics in the generated simulations to their high-resolution counterparts. 
  We find that the model accurately reproduces large-scale statistics, robustly recovering most of the power that was missing from the low-resolution input, but exhibits a residual suppression of power on small scales of up to $\sim 10\%$ at $k \sim 1\,h\,\mathrm{Mpc}^{-1}$. The abundance of halos around $10^{14}\,M_\odot$ is affected at a similar level, and we find that the profiles of these halos have a lower central density. Although the overall performance is decent, we anticipate that the fidelity of the generative model can be further increased with more and better training data, as well as through improvements in the model architecture and training process. To show a production-scale use case, we apply our model to upscale the resolution of a light cone from a large-volume $N$-body simulation with spatial curvature, producing a first-of-its-kind catalogue that simultaneously captures geometric effects at large scales and accurate nonlinear structure at small scales.
  }

   \keywords{large-scale structure of Universe - methods: numerical}

   \titlerunning{}
   \authorrunning{Fremstad et al.}
   \maketitle

\section{Introduction}
\label{sec:intro}

The current generation of Stage-IV galaxy surveys such as DESI \citep{DESI:2013agm}, Euclid \citep{Euclid:2024yrr}, or the Vera C.\ Rubin Observatory's Legacy Survey of Space and Time \citep[LSST,][]{LSST:2008ijt}, and their upcoming radio counterparts such as HIRAX \citep{Newburgh:2016mwi} or SKA1-MID \citep{SKA:2018ckk}, probe an enormous volume of the observable universe. This allows us to measure correlations over an unprecedented range of scales, from sub-Mpc to several Gpc. The statistical power of these surveys also requires very detailed and precise forward models to interpret the data, often based on end-to-end numerical simulations that connect a cosmological model with observed summary statistics. In these forward models, the gravitational dynamics are typically modelled with $N$-body simulations that can be used as a reference to compare directly to the data, as a way to estimate covariances, or as the `ground truth' for training emulators. $N$-body simulations that simultaneously cover multi-Gpc correlations and sub-Mpc detail pose a major computational challenge due to the large number of mass elements required. For example, the Euclid Flagship 2 simulation \citep{Euclid:2024few} has four trillion mass elements.

Machine learning offers a pathway to generating such large synthetic datasets in a more economic way. An idea would be to run the large-volume $N$-body simulation at much lower resolution; this is relatively cheap and provides an accurate model for the fluctuations at ultra-large scales. Separately, we can train a machine-learning algorithm to learn the distribution of high-resolution data conditioned on low-resolution context and relevant physical parameters (e.g.\ cosmology). Pairs of high-resolution, small-volume simulations and low-resolution versions can be used as training examples. That way we can build a `Super-Resolution Emulator' that can generate highly realistic small-scale detail for any given low-resolution context. By using state-of-the-art high-fidelity generative models, such as generative adversarial networks \citep[GANs,][]{Zhang:2023lqi, Ramanah:2020vyl, Sipp:2022tdp} or diffusion models \citep{Rouhiainen:2023ewv}, such emulators are able to reconstruct high-resolution features with good accuracy, usually only inducing a small error around $\lesssim 10\%$ at the very smallest scales.

In this paper, we explore this approach in a particularly relevant use-case scenario: the generation of synthetic large-scale structure data for non-vanishing spatial curvature. Although the current standard model of cosmology, the Lambda-cold-dark-matter model ($\Lambda$CDM), assumes a spatially flat geometry, this assumption is routinely tested with cosmological observations. While being relatively straightforward to accommodate theoretically, a conclusive detection of curvature would have huge implications, not least because it would completely overturn the currently prevailing paradigm of cosmic inflation. Current constraints are often based on analyses that use background cosmology plus linear perturbation theory, such as the ones of Planck \citep{Planck:2018vyg} or DESI \citep{DESI:2025zgx}, and are typically of the order of a few percent, depending on model assumptions. However, it has also been argued that a curvature value as small as $0.2\%$ can have important implications for precision measurements in cosmology, such as the one of the absolute neutrino mass scale \citep{Chen:2025mlf}. In the future, we may therefore want to use the full non-linear information contained in the large-scale structure to derive additional constraints and thus reduce overall model dependence.

The presence of curvature would affect our observations of large-scale structure in two important ways. First, it changes the expansion history and hence the growth rate of structure. This effect is present on all scales and therefore affects also structures much smaller than the curvature scale for which the geometry looks essentially Euclidean. Second, there is a geometric effect on how we observe structure over large distances down our past light cone. This effect does depend on the scales probed, and hence becomes more relevant as survey volumes increase. \citet{DiDio:2016ykq} have conducted a detailed study in linear perturbation theory that disentangles the two effects, and they found that the geometric effect can actually be more important than the effect on the growth rate when considering angular clustering in wide redshift bins typical for photometric surveys or weak lensing tomography. Furthermore, in scenarios with evolving dark energy like those favoured by recent DESI results \citep{DESI:2025zgx}, the effect of curvature on the expansion history (and hence the growth rate) is completely degenerate with dark energy, and disentangling the two requires a true geometric probe of non-Euclidean geometry.

The modified expansion history under curvature has traditionally been taken into account in $N$-body simulations through the separate universe approximation \citep{Li:2014sga,Wagner:2014aka}. Here, the simulation box is considered much smaller than the curvature scale so that the geometry inside the box can be treated as Euclidean. This leaves the open question of how such separate universe boxes should be stitched together consistently to obtain a large-scale survey footprint that is sensitive to the geometric effect of curvature. For such large scales, the problem can be solved through the approach developed by \citet{Adamek:2025ylh}. The idea here is to effectively simulate a Lema\^itre-Tolman-Bondi model, i.e.\ a model with a radial curvature profile. The profile is chosen non-zero and constant in a large patch that contains the survey footprint and is apodized outside the light cone to accommodate periodic boundary conditions. While this approach captures the geometric effect of curvature perfectly, it requires the simulation volume to be large enough to fit the entire survey footprint.

Here we present a methodology that combines the two approaches using machine learning. We train a Super-Resolution Emulator on a set of separate universe simulations, and then apply it on a large-volume, low-resolution particle light cone to recover the small-scale detail in a probabilistic sense. The remainder of the paper is structured as follows. In \cref{sec:prelim}, to set the stage we briefly review some basic concepts. Our methodology for training the emulator is described in \cref{sec:methods}, and we evaluate the performance of the trained model in \cref{sec:perf} on test data using summary statistics. In \cref{sec:bigsim}, we apply the Super-Resolution Emulator to a production-scale simulation, generating a first-of-its-kind particle light cone for a curved universe model. We conclude in \cref{sec:fin}.

\section{Preliminaries}
\label{sec:prelim}

\subsection{The Friedmann universe with curvature}

A homogeneous universe with spatial curvature is characterised by the well-known Friedmann-Lema\^itre-Robertson-Walker (FLRW) line element
\begin{equation}
    ds^2 = -dt^2 + a^2(t) \left[\frac{dr^2}{1-k r^2} + r^2 d\Omega^2\right]\,,
\end{equation}
where $a(t)$ is the scale factor, $r$ is a comoving radial coordinate related to the circumference (or area) of a sphere around the origin, and $k$ is the Gaussian curvature of space at $a = 1$. When $k \neq 0$, the usual Euclidean relation between diameter and circumference is distorted: for a sphere of given circumference, its diameter for positive (negative) curvature is larger (smaller) than usual. The amount of distortion depends on the ratio between the size of the sphere and the curvature scale.

Inserting this ansatz into Einstein's equations yields the Friedmann equation
\begin{equation} \label{eq:Friedmann}
    H^2 = \frac{\dot a^2}{a^2} = \frac{8 \pi G}{3} \rho + \frac{\Lambda}{3} - \frac{k}{a^2}\,,
\end{equation}
which fixes the time evolution of the scale factor. As noted earlier, the presence of curvature has a distinct effect on the expansion history, acting similar to a matter source with equation of state $w = -^1\!/\!_3$. While the expansion history affects the growth of structure on all scales, there is a complete degeneracy with evolving dark energy. That is, if the equation of state of dark energy is left completely unspecified so that the cosmological constant $\Lambda$ in \cref{eq:Friedmann} is replaced by an arbitrary function of time, there is no possibility to constrain curvature separately by only looking at the expansion history.

To disentangle curvature from evolving dark energy, we may look at geometric probes. Spatial curvature directly affects distance measures in cosmology. For example, the angular diameter distance is given by
\begin{equation}
    d_A(z) = (1+z)^{-1} \begin{cases} 
          k^{-1/2} \sin\left(k^{1/2} \int\limits_0^z H^{-1}(z') dz'\right) & k > 0\\
          \int\limits_0^z H^{-1}(z') dz' & k = 0 \\
          \vert k\vert^{-1/2} \sinh\left(\vert k\vert^{1/2} \int\limits_0^z H^{-1}(z') dz'\right) & k < 0\,,
       \end{cases}
\end{equation}
where $z$ is the redshift of a source in the FLRW model. If we look at the Taylor expansion for small $z$, we note that the curvature parameter $k$ changes the distance as $\Delta d_A / d_A \approx -k z^2 / (6 H_0^2) \approx -k d_A^2 / 6$. This highlights the fact that one needs a sufficiently long baseline to measure curvature in this way.

\subsection{Generative adversarial networks}

To address the difficulty in approximating many intractable probabilistic computations arising in maximum likelihood estimation, \cite{Goodfellow:2014upx} proposed a framework known as generative adversarial networks (GANs), which completely avoids explicit likelihood computation. In this framework, a strategy is employed involving two competing networks. The {\it generator} ($G$) network is tasked with generating fake samples, while the {\it discriminator} ($D$) is tasked with distinguishing between real samples and fake ones generated by the generator. 
By training the generator to maximise the misclassification of the discriminator, it learns $p_g$, an approximate mapping from an input noise variable ${\bf x} \sim p_{\bf x}$ to the true data distribution $p_{\bf y}$. 
% The discriminator is trained to map input samples to $[0,1]$, where a value of $1$ implies that the discriminator has labelled the sample as real. 

Conditional GANs (cGANs) introduce additional information, ${\bf z} \sim p_{\bf z}$, which serves to condition the generation and discrimination process. The objective functions of the generator ($\mathcal{L}_G$) and the discriminator ($\mathcal{L}_D$) may thus be written as
\begin{align}
    \mathcal{L}_D &= +\mathbb{E}_{\bf x}\left[D(G({\bf x}|{\bf z})|{\bf z}) \right] - \mathbb{E}_{\bf y} \left[D({\bf y}|{\bf z}) \right] \,,\\
    \mathcal{L}_G &= -\mathbb{E}_{\bf x}\left[ D(G({\bf x}|{\bf z})|{\bf z}) \right]\,.
\end{align}

This framework is notoriously delicate and unstable. Without careful regularisation or architectural choices, GANs may suffer from mode collapse, where the generator fails to capture the diversity of the data distribution and instead settles on producing a limited subset of outputs that successfully fool the discriminator without actually representing the true data distribution. Furthermore, when the real and generated distributions have disjoint support, the discriminator can become overly confident, resulting in vanishing gradients for the generator and stalled training.

Wasserstein GANs (WGANs) were proposed by \citet{Arjovsky:2017fjr} to address these issues. Rather than minimising the Jensen-Shannon divergence, WGANs approximate the Wasserstein-1 distance, or the Earth-Mover distance, between the real and generated distributions using the Kantorovich-Rubinstein dual formula. This requires that the discriminator be Lipschitz continuous with a Lipschitz constant of 1, which corresponds to the constraint $|D(x_1) - D(x_2)| \leq |x_1 - x_2|$. To maintain this constraint, \citet{Gulrajani:2017kpy} proposed adding a gradient penalty regularisation term to the loss function. The final loss may be written as 
\begin{equation}
    \mathcal{L}_D = +\mathbb{E}_{\bf x}\left[D(G({\bf x}|{\bf z})|{\bf z}) \right] - \mathbb{E}_{\bf y} \left[D({\bf y}|{\bf z}) \right] + \lambda \mathbb{E}_{\bf x} \left[\left(||\nabla_{\bf x} D({\bf x}|{\bf z})||_2 - 1 \right)^2 \right],
\end{equation}
where $\lambda$ is a hyperparameter to control the magnitude of the gradient penalty. For computational efficiency, we apply the gradient penalty once every 8 batches. 
\section{Methods}
\label{sec:methods}

\subsection{$N$-body simulations with curved geometry}

To take into account the full effect of non-Euclidean geometry, \citet{Adamek:2025ylh} construct a spacetime solution given by a homogeneous ball of dust that freely expands into vacuum. The initial density and expansion rate of the ball can be chosen such that the interior of the ball is a patch of a closed (or open) FLRW solution that shall contain the survey footprint of the simulated observations. To arrange periodic boundaries, which are the most convenient choice for $N$-body simulations, the surrounding vacuum region is itself embedded as a cavity in a flat FLRW dust model. The three different regions, curved FLRW, vacuum (Schwarzschild-de\,Sitter), and flat FLRW, are arranged in such a way that the entire spacetime remains a faithful solution to Einstein's equations. Observables are computed consistently by ray tracing the light cone inside the curved patch.

The drawback of this approach, which we address in this work, is the requirement that the entire light cone relevant to the simulated observation has to fit inside the curved patch, as the vacuum region and flat exterior are unphysical constructs to implement the correct boundary conditions. Observables that can constrain curvature would typically cover large volumes and extend to high redshifts, as this is where geometric effects would be most important. With limited memory and compute resources at our disposal, this usually implies that such $N$-body simulations will have to be run at relatively modest mass resolution. For example, the simulations presented in Sec.~\ref{sec:bigsim} have a mass resolution of about $6.8 \times 10^{12}\,M_\odot$. Furthermore, the mesh used in our particle-mesh code has a similarly modest resolution of about $4\,h^{-1}\,\mathrm{Mpc}$. While this is sufficient to probe structure formation into the nonlinear regime (the density contrast crosses unity on scales larger than the resolution limit), the ability to capture halo formation is significantly curtailed.

Having constructed a coarse view of the survey footprint that accurately accounts for large-scale correlations, including geometric effects, our aim in this paper is to augment this view by filling in small-scale detail that is physically realistic and consistent with the coarse-grained perspective. To this end, we return to the separate universe approach, i.e.\ the simulation of structure formation in small (periodic) boxes that follow the expansion law of a curved FLRW model but can be considered Euclidean on the scale of the box. We train a super-resolution emulator on pairs of such simulations, where each pair consists of a low-resolution input and a high-resolution target that shares the same initial data on the low-resolution scales.

\subsection{Training data}

Our low-resolution inputs are tuned to have the same spatial resolution as our large-volume light-cone run, i.e.\ about $4\,h^{-1}\,\mathrm{Mpc}$. These simulations are extremely cheap because they have a small volume and low resolution, and they are run with \texttt{gevolution} \citep{Adamek:2015eda,Adamek:2016zes} which uses a particle-mesh algorithm at fixed mesh resolution. The high-resolution targets are run with eight times more particles using \texttt{Gadget-4} \citep{Springel:2020plp}, increasing the force resolution by orders of magnitude compared to the fixed-grid input. Specifically, we use a gravitational softening length of $50\,h^{-1}\,\mathrm{kpc}$. The flexibility to use two different codes where one of them is better adapted to capture small-scale physics is a huge advantage of our machine-learning framework. Our augmentation process is not restricted to emulating what a higher-resolution simulation of the same code would produce, but it can simultaneously learn to correct for missing capabilities of a simulation framework, like the absence of adaptive force resolution and time stepping.

\begin{figure}
    \centering
    \includegraphics[width=\linewidth]{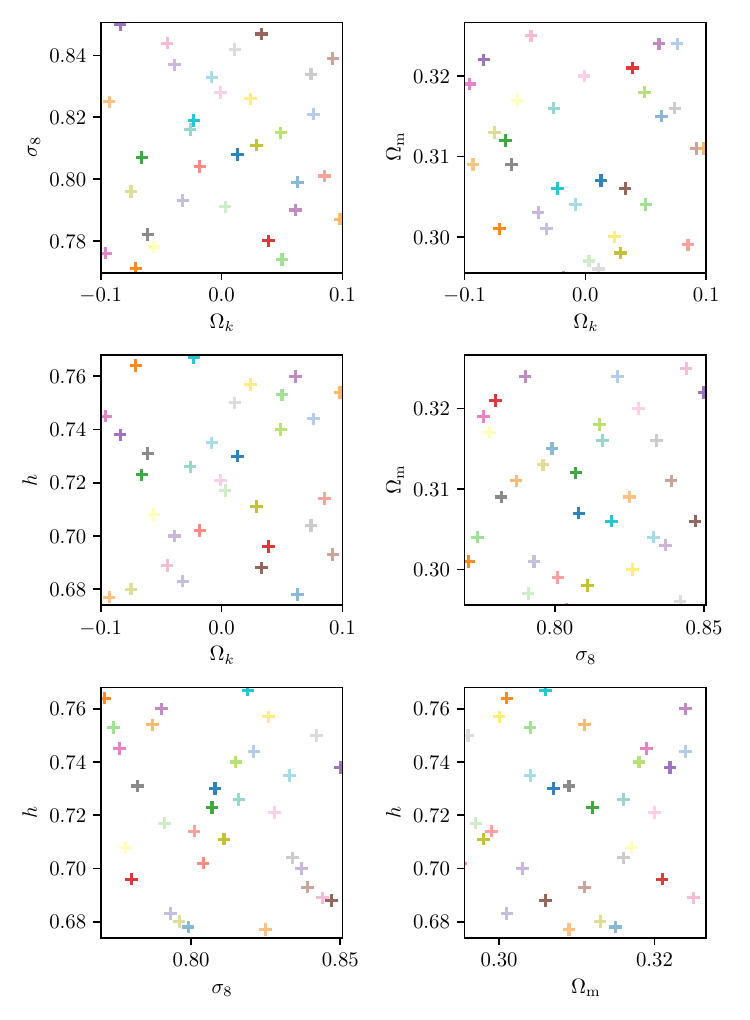}
    \caption{Projection of the Latin hypercube sampling into the different two-dimensional parameter ranges. We sample uniformly $\Omega_k \in \left[-0.1, 0.1\right]$, $h \in \left[0.674, 0.768\right]$, $\sigma_8 \in \left[0.76969, 0.85071\right]$, and $\Omega_\mathrm{m} \in \left[0.29555, 0.32666\right]$. The different colours correspond to the 32 different training examples.}
    \label{fig:latin-hypercube}
\end{figure}

As explained in more detail below, our generative model is conditioned on four cosmological parameters, $\Omega_k$, $\Omega_\mathrm{m}$, $\sigma_8$ and $h$, as well as on redshift to capture time dependence of large-scale structure. The effect of $\Omega_k$ is modelled here using the separate universe approach, i.e.\ it enters only via the Friedmann equation \eqref{eq:Friedmann} for the background. Our training data sample the four-dimensional cosmological parameter space using a Latin hypercube with 32 sampling points. To achieve an even coverage of the parameter space, we enforce that each orthant (hyperoctant) of the four-dimensional Cartesian product of parameter ranges contains exactly two samples ($2^4\times 2 = 32$). Our sampling of cosmological parameters is illustrated in Fig.~\ref{fig:latin-hypercube}. Time evolution is sampled using 11 snapshots, spaced uniformly in the value of the scale factor, $a \in \left\{0.25, 0.3, \ldots, 0.75\right\}$.

To automate the process of generating settings files and run scripts, we repurpose and adapt a pipeline published with the \textit{Quijote} simulations \citep{Villaescusa-Navarro:2019bje}. Each simulation pair consists of a low-resolution box with $L_\mathrm{box} = 1000\,h^{-1}\,\mathrm{Mpc}$ and $256^3$ particles and a high-resolution counterpart with $512^3$ particles. As mentioned earlier, each pair uses the same initial data: we first generate the displacement field for the high-resolution simulation using second-order Lagrangian perturbation theory; the low-resolution initial data are then obtained through deterministic downsampling. However, to avoid training specific random patterns into the model, we vary the random realisation when sampling the Latin hypercube, i.e.\ we use 32 different random seeds.

The linear matter power spectrum that enters the Lagrangian perturbation theory is computed with \texttt{CAMB} \citep{Lewis:1999bs}, and apart from the cosmological parameters already discussed, we fix the baryon density $\Omega_\mathrm{b} = 0.049$, the photon temperature $T_\mathrm{CMB} = 2.725\,\mathrm{K}$, and the scalar spectral index $n_\mathrm{s} = 0.9624$, and we assume three massless neutrino species with the standard value $N_\mathrm{eff} = 3.046$. The linear matter power spectrum at redshift $z=0$ is scaled back to the initial redshift $z_\mathrm{in} = 63$ using the Newtonian growth factor in a universe without radiation so that radiation can then be neglected for simplicity.

We represent the training data using displacement fields as it offers a discrete representation of the simulation snapshots that, by construction, falls on a regular grid. 
This formulation is well suited for convolutional neural networks, which operate efficiently on grid-based data through convolutional kernels. Training on displacement fields also has the benefit that one can easily calculate the particle positions. 
Specifically,
\begin{equation}\label{eq:displacement}
    \boldsymbol{x} = \boldsymbol{q} + \boldsymbol{\Psi}(\boldsymbol{q}, t)\,,
\end{equation}
where $\boldsymbol{q}$ is the unperturbed initial position of the particle, $t$ is cosmic time, $\boldsymbol{\Psi}$ is the displacement vector, and $\boldsymbol{x}$ is the final position at time $t$. 

Following \citet{Li:2020vor}, we use cropping to reduce the memory load of a training step. Low-resolution boxes are cropped into sub-boxes with $L_{\rm box} = 125 h^{-1}{\rm Mpc}$, which corresponds to a displacement field with $3\times 64^3$ pixels. The high-resolution boxes are cropped to $2\times$ the resolution of the low-resolution fields, i.e., the displacement field with $3\times128^3$ pixels. In addition, data augmentation is utilised to extend the data-set through random rotations of the sub-boxes, which also enforces symmetries in the training process. Furthermore, padding of the low-resolution sub-boxes is employed to provide context for pixels along the edges. 

\subsection{Model architecture}

For a given low-resolution (LR) simulation patch $\boldsymbol{\Psi}_\mathrm{LR}$ and cosmological parameters $\phi$, the goal of the model is to learn the mapping 
\begin{equation}
    \mathcal{F}_\theta : \left\{\boldsymbol{\Psi}_\mathrm{LR}(\boldsymbol{q}), \phi\right\} \rightarrow \boldsymbol{\Psi}_\mathrm{HR}(\boldsymbol{q})\,,
\end{equation}
where $\boldsymbol{\Psi}_\mathrm{HR}$ is the high-resolution (HR) counterpart to the given input. The LR patch lacks small-scale (high-$k$) information due to finite force and mass resolution, which must be recovered by the model. 
Due to cosmic variance, these HR counterparts are not uniquely defined, so there exists an ensemble of physically valid HR realisations for a given LR patch and cosmological parameters. To emulate this, one can choose a stochastic approach to recover missing modes rather than a deterministic one. 

The reconstruction of the model comes in two stages. First, the model naively upsamples the LR input via interpolation. Then, the model predicts the residual field, which captures unresolved small-scale structure. Through this formulation, the model capacity will be focused on missing information rather than reproducing large-scale modes, which will be preserved by construction. 
We employ a U-net encoder-decoder architecture to predict the residual field, where the encoder extracts multi-scale contextual information, while the decoder reconstructs HR structure. Our model is completely convolution-based, allowing for patch-based processing, which enables scalability to larger cosmological volumes. 

To introduce stochasticity and high-frequency features into the model, noise is intermittently added to the feature maps, following the approach by \citet{Zhang:2023lqi}. This allows for the generation of diverse HR realisations from the same LR patch. 
To successfully incorporate the noise as high frequency structure, adversarial training is required, which is done by introducing a discriminator network that operates on statistical properties rather than pixel-wise accuracy. This shifts the learning task from purely reproducing the target field to generating a field that is physically consistent with the target, ensuring agreement with the target simulation ensemble. 

Our model is conditioned on the scale factor, $\Omega_k$, $\sigma_8$, $\Omega_\mathrm{m}$, and $h$. Conditioning is implemented through modulated convolution, styled noise, and styled activations. 
Modulated convolution works by passing the conditioning (or style), which in our case will be the cosmological parameters and the scale factor, through a neural network, which transforms the conditioning into a latent representation, which modifies the convolutional kernels so as to dynamically adapt the feature extraction process to reflect the influence of those parameters on the generated output. 
Similarly, we are able to pass the cosmological parameters through a neural network to produce latent representations, which determine the mean and standard deviation of the noise added to the feature maps, and the slope of the ReLU activation function. This is especially useful, as the amplitude of the added noise can be modified to reflect the amount of non-linear structure present at a given redshift. 

\section{Model Performance}
\label{sec:perf}

To evaluate the accuracy of our model, we set aside a small part of the data, known as test data, which the model does not encounter during training. 
This allows us to determine how well our model does at interpolating between known samples, revealing how well it has fitted to the data. 

Through \cref{eq:displacement}, it is possible to construct the position of particles by adding the predicted super-resolution (SR) displacement field $\boldsymbol{\Psi}_\mathrm{SR}$ to the initial particle positions $\boldsymbol{q}$, which represent the unperturbed initial conditions. Given the reconstructed particle positions, we can use standard cosmological probes to assess the realism of the generated samples by comparing the model prediction to known high-resolution counterparts. 
Since our model does not map the LR input to the specific HR counterpart provided in the data but rather to an ensemble of realistic HR realisations, the fidelity of the model should be assessed using summary statistics, which is common for cosmological probes. 

In \cref{sec:bigsim}, to show a possible use case, we employ our model to upscale a large simulation of a curved universe. In this case, although we do not have a HR counterpart to compare, we are still able to investigate the model's performance by comparing it to independent predictions of the angular correlation function.

\subsection{Matter power-spectrum}

The matter power spectrum describes the variance of density fluctuations as a function of scale, which we can use as a metric to measure discrepancies in predicted structure when compared to the ground truth. 
The matter power spectrum is the Fourier transform of the 2-point correlation function $\xi(r)$, defined as 

\begin{align}
    \xi(|\boldsymbol{r}|) &= \langle \delta(\boldsymbol{x}) \delta(\boldsymbol{x} + \boldsymbol{r}) \rangle\,, \\
    P(k) &= (2 \pi)^{-3}\int \xi(|\boldsymbol{r}|) \,e^{-i \boldsymbol{k} \cdot \boldsymbol{r}} d^3 \boldsymbol{r}\,,  
\end{align}
where $\delta(\boldsymbol{x}) = \frac{\rho(\boldsymbol{x})}{\bar{\rho}} - 1$ is the overdensity, and $\boldsymbol{k}$ is the 3D wave vector of the plane wave. 

\begin{figure}
    \centering
    \includegraphics[width=\linewidth, trim=0 6mm 18mm 20mm, clip]{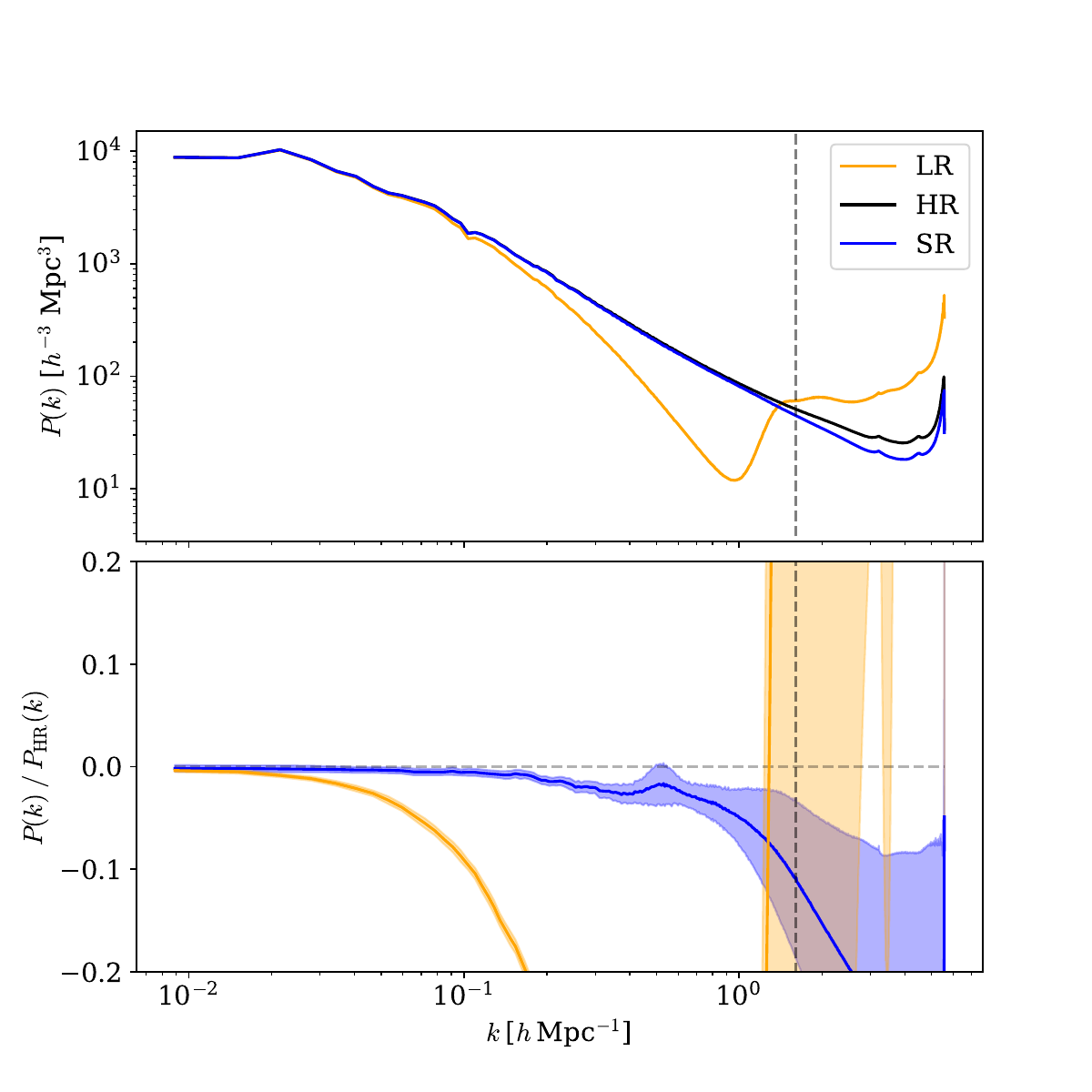}
    \caption{Comparison of the matter power spectrum at $z \approx 0.817$ of the LR input, the HR target, and the SR model prediction. The bottom panel shows the mean and standard deviation (bands) in the relative difference between the predicted and the target power spectrum of the whole test data-set. The vertical dashed line indicates the Nyquist wavenumber of the high-resolution initial data. To estimate the spectra, we project the density onto a $2\times$ finer grid so that the estimator has very small discretisation effects at the original Nyquist scale.}
    \label{fig:test-set-accuracy} 
\end{figure}

We compute the SR fields for all LR samples in the test data-set. The mean and scatter of the relative difference between the power spectra of the SR and corresponding HR samples are shown in the bottom panel of \cref{fig:test-set-accuracy}. To compute power spectra, we use the Python package \texttt{Pylians3} \citep{Pylians}. We find that we successfully recover the HR spectra on large scales $k \lesssim 0.2\,h\,\mathrm{Mpc}^{-1}$ within $\sim 1\%$. On smaller scales, the spectra match within $\sim 5\%$ for $k \lesssim 1\,h\,\mathrm{Mpc}^{-1}$, and within $\sim 10\%$ up to the Nyquist wavenumber.
Since baryonic effects become significant at scales smaller than $k\sim 1\,h\,\mathrm{Mpc}^{-1}$ \citep{Chisari:2018prw, Huang:2018wpy}, we cannot accurately model the dark matter power spectrum at these scales without a full hydrodynamic implementation. Thus, we do not require that our model is able to accurately reproduce the dark matter distribution at these scales.

\begin{figure}
    \centering
    \includegraphics[width=\linewidth, trim=0 6mm 18mm 20mm, clip]{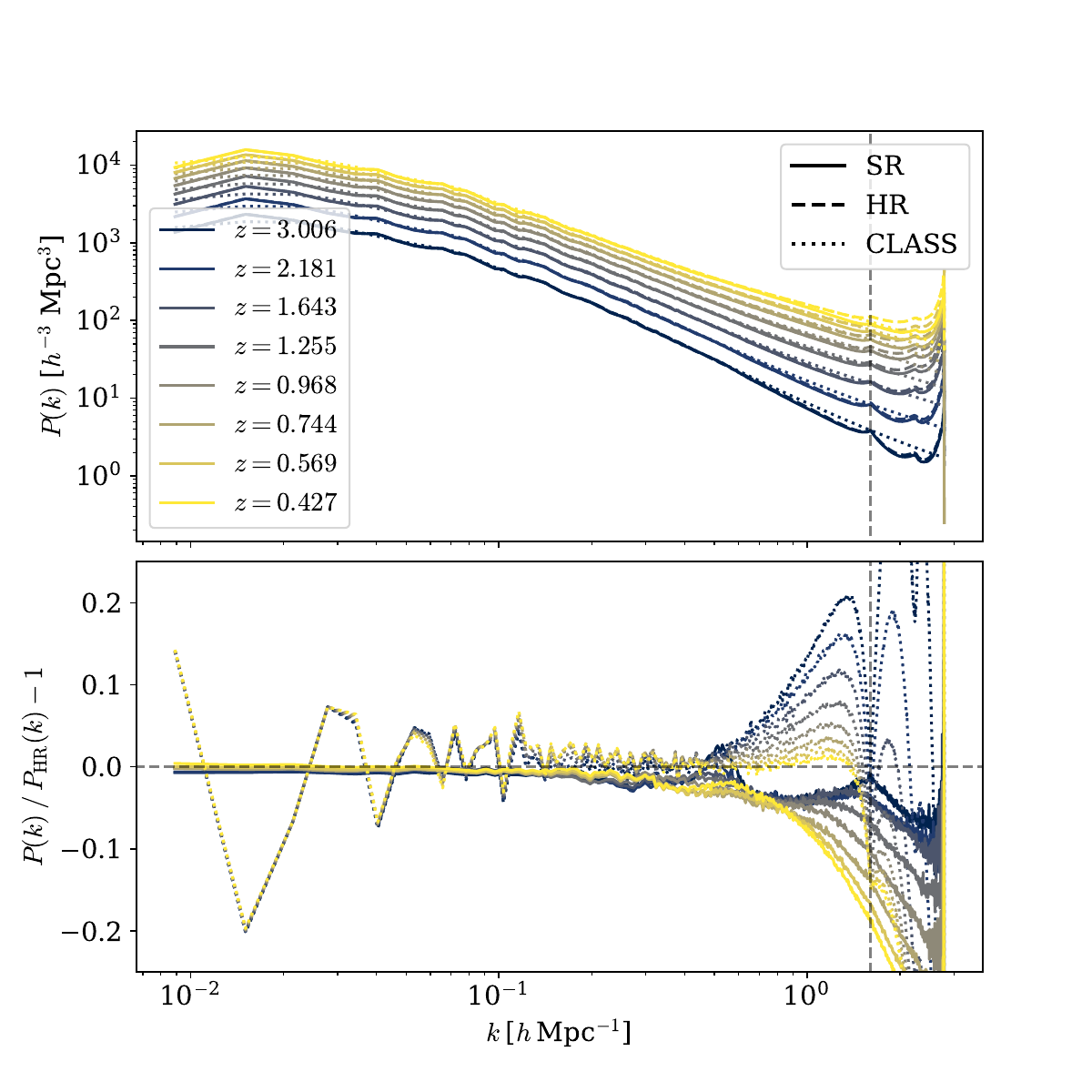}
    \caption{Matter power spectrum of SR samples, along with their HR counterpart and the prediction from \texttt{CLASS} with the nonlinear model published by \citet{Mead:2020vgs} as a function of redshift. Here, we have used a finer sampling over redshift than what was used in the training data in order to also test the model's ability to interpolate. The relative error with respect to the HR target is shown in the lower panel.
    }
    \label{fig:redshift-accuracy}
\end{figure}

In \cref{fig:redshift-accuracy}, we compare the predicted matter power-spectrum to the corresponding HR reference across a range of redshifts. To evaluate the model's ability to interpolate over time, we run a simulation from the training dataset with finer sampling in redshift, thereby probing redshift values not explicitly seen during training. Across all redshifts, the model is able to successfully reproduce the matter power spectrum, only struggling on small scales. At $k \sim 1\,h\,\mathrm{Mpc}^{-1}$,
the predicted matter power spectrum agrees with the target within
$\sim 10\%$, with the largest deviations observed at the lowest redshifts.

The error seen in \cref{fig:redshift-accuracy} can likely be attributed to two factors. First, the prediction task becomes easier at high redshift because structure formation is still largely linear. As a result, the LR input field and the corresponding HR field are more similar, meaning that the residual structure the model attempts to reconstruct is less complex. Consequently, the model is able to match the target power spectrum more accurately at high redshift, while larger deviations appear at lower redshift, where nonlinear structure becomes more prominent. 

Second, our model is trained across a wide range of redshifts and must therefore learn a mapping that generalises across different stages of structure formation. This increases the complexity of the learning problem and may lead to reduced accuracy in regimes where the dynamics are more difficult to model. Increasing the expressive power of the conditioning mechanism could help alleviate this limitation. In our case, conditioning is implemented through modulated convolution, but more flexible approaches -- perhaps such as attention-based conditioning or transformer architectures -- may allow the model to more effectively capture the evolving statistical properties of the field across cosmic time. An alternative strategy is to fine-tune the model to specific redshift ranges, which can simplify the learning task by having a specific model for linear regimes, and one for non-linear regimes. This is the approach taken in \citet{Zhang:2023lqi}.

\subsection{Velocity reconstruction}\label{sec:velocity}

Although our model does not predict the velocity of particles directly, they can be reconstructed using a finite-difference approximation. By upscaling two displacement fields separated by some time interval $dt$, the peculiar velocity can be estimated as
\begin{equation}
    \boldsymbol{v} = a \frac{d\boldsymbol{x}}{dt}\,,
\end{equation}
where $d\boldsymbol{x}$ is obtained from the difference between the two reconstructed displacement fields. This is done by advancing the particle positions of a LR snapshot at some redshift $z$ by some interval, $\pm dz$, according to the equation 
\begin{equation}
    \boldsymbol{x}(z \pm dz) = \boldsymbol{x}(z) \mp \frac{\boldsymbol{v} dz}{H_0 \sqrt{\Omega_m a^{-3} + \Omega_\Lambda + \Omega_k a^{-2}}}\,.
\end{equation}
By upscaling the particle positions, we estimate $d\boldsymbol{x}$ through
\begin{equation}
    d\boldsymbol{x} = G(\boldsymbol{x}(z-dz)) - G(\boldsymbol{x}(z+dz))\,,
\end{equation}
where $G$ is the generator model. 

Since our model generates SR fields stochastically, each output represents a sample from an ensemble of possible HR realisations. To ensure that the two reconstructed fields correspond to the same underlying structure, the upscaling must therefore be performed using the same noise realisation in both cases. This ensures that the differences between the fields arise only from the time evolution of the system, rather than from stochastic variations in the model output.

\begin{figure}
    \centering
    \includegraphics[width=\linewidth, trim=0 6mm 18mm 20mm, clip]{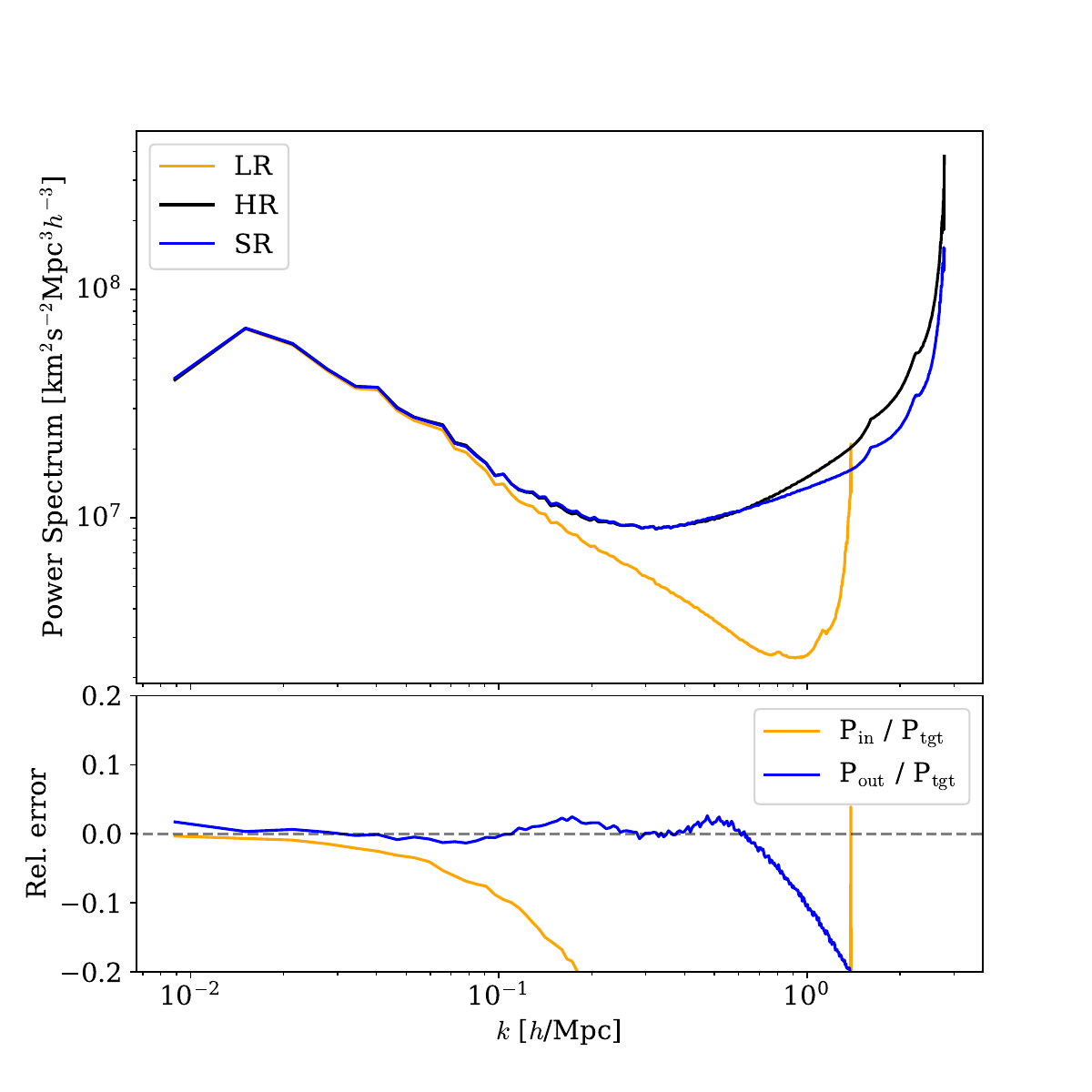}
    \caption{Power spectrum of the divergence of the velocity field from the LR and HR simulations, along with the reconstruction from the SR model. The power spectrum is calculated at $z=0.7$. To ensure consistency, all spectra are calculated using the reconstruction method described in \cref{sec:velocity} using $dz=10^{-5}$.}
    \label{fig:velocity-pk}
\end{figure}

The accuracy of the reconstructed velocity field can be assessed by comparing the calculated velocity power spectrum with that of the HR simulations. We present such a comparison in \cref{fig:velocity-pk}, where we find that we are able to extend the range of accurately resolved velocities significantly compared to the LR input. At large scales $k \lesssim 0.7$, our reconstructed velocity agrees well with the HR target, only inducing an error of $\lesssim2\%$. For smaller scales, as seen in the previous section, our model deviates more strongly from the target, resulting in an error of around $\sim 10\%$ at $k \sim 1\,h\,\mathrm{Mpc}^{-1}$.

The velocity reconstruction might suffer due to the fact that our model is  trained on individual snapshots rather than on the full time-evolution of the system. 
As a result, dynamical consistency between consecutive time-steps is not enforced, and there is no guarantee that structures evolve coherently between snapshots. Since the velocity field is obtained from differences between reconstructed displacement fields, it is particularly sensitive to small, uncorrelated errors in the predicted particle positions. In addition, velocities depend strongly on small-scale gradients, making them especially susceptible to the inaccuracies observed at small scales in \cref{fig:redshift-accuracy}. Improving the reconstruction of the velocity field is left for future work. In \cite{Ni:2021mzk}, this is addressed by training the model to jointly predict the displacement and velocity fields, enforcing a more consistent dynamical representation. Alternatively, one could train a separate model dedicated to predicting velocities from the reconstructed particle distribution.

\subsection{Halo mass function and halo profiles}

Halo properties can offer further insight into the accuracy and realism of our model at small scales, as dark matter halos trace the nonlinear regime of structure formation. 
We identify halos using the {\tt yt} framework \citep{yt} with the {\tt yt\_astro\_analysis} extension \citep{yt.astro.analysis}, 
running the friends-of-friends algorithm \citep{Efstathiou:1985re} with a linking length of $b=0.2$. 
From the identified halos, we extract the halo mass function and radial density profiles. These quantities probe both the abundance of collapsed structure and their internal morphology, providing a stringent test of the model beyond two-point statistics. 

\begin{figure}
    \centering
    \includegraphics[width=\linewidth, trim=0 6mm 18mm 20mm, clip]{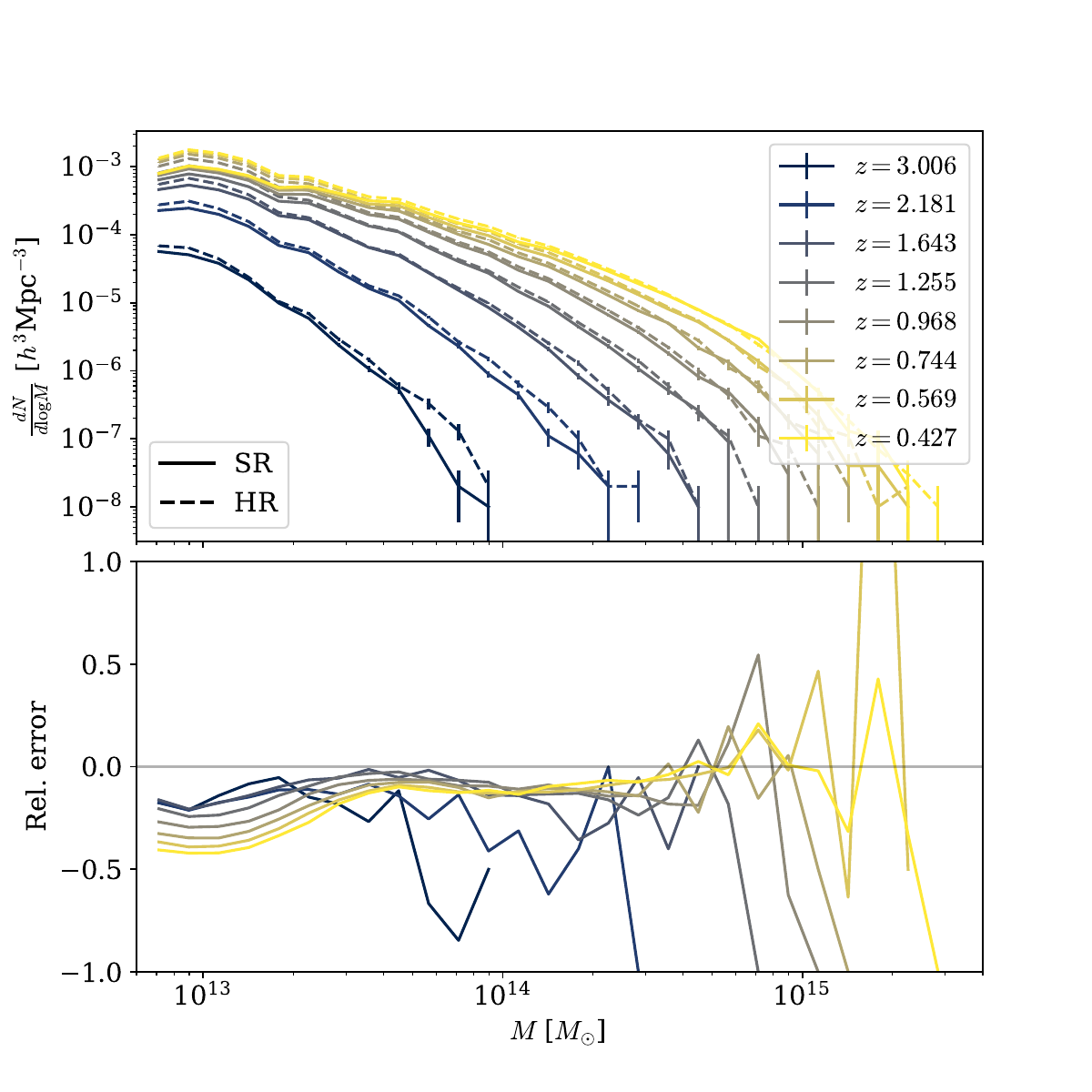}
    \caption{The halo mass function at different redshifts. The upper panel shows the comparison between the produced SR, and the simulated HR halo mass functions. The lower panel shows the relative difference between the two. }
    \label{fig:HMF}
\end{figure}

We compare the halo mass function between the generated SR particles and the HR ones over a range of redshifts in \cref{fig:HMF}. The bottom panel shows the relative difference between the two. Overall, the SR prediction contains $\sim 20\%$ fewer halos at high redshift, growing to a loss of about $\sim 30\%$ of halos at low redshift. We observe a reduction in the number density of low-mass halos of $\lesssim 40\%$, while for intermediate-mass halos (around $5\times 10^{13}\,M_\odot$), the reduction reaches $\sim 10\%$.  

The increased error at small scales observed in \cref{fig:redshift-accuracy} likely contributes to the reduced number density towards the low-mass end of the halo mass function. Low-mass halos typically contain only a few particles, making them highly sensitive to small-scale errors. Even modest perturbations in the particle positions can therefore prevent marginal halos from forming or cause them to fall below the detection threshold of the halo finder.
These errors may also slightly affect halos of somewhat higher mass, as misplaced particles can alter halo membership and shift halos between neighbouring mass bins. However, this effect is expected to be weaker for more massive halos, which contain many particles and are therefore more robust to small perturbations in particle positions. Indeed, for halo masses above $\sim 5\times 10^{14}\,M_\odot$, the halo mass function of the SR realisations is statistically consistent with the HR target.

\begin{figure}
    \centering
    \includegraphics[width=\linewidth, trim=0 6mm 18mm 20mm, clip]{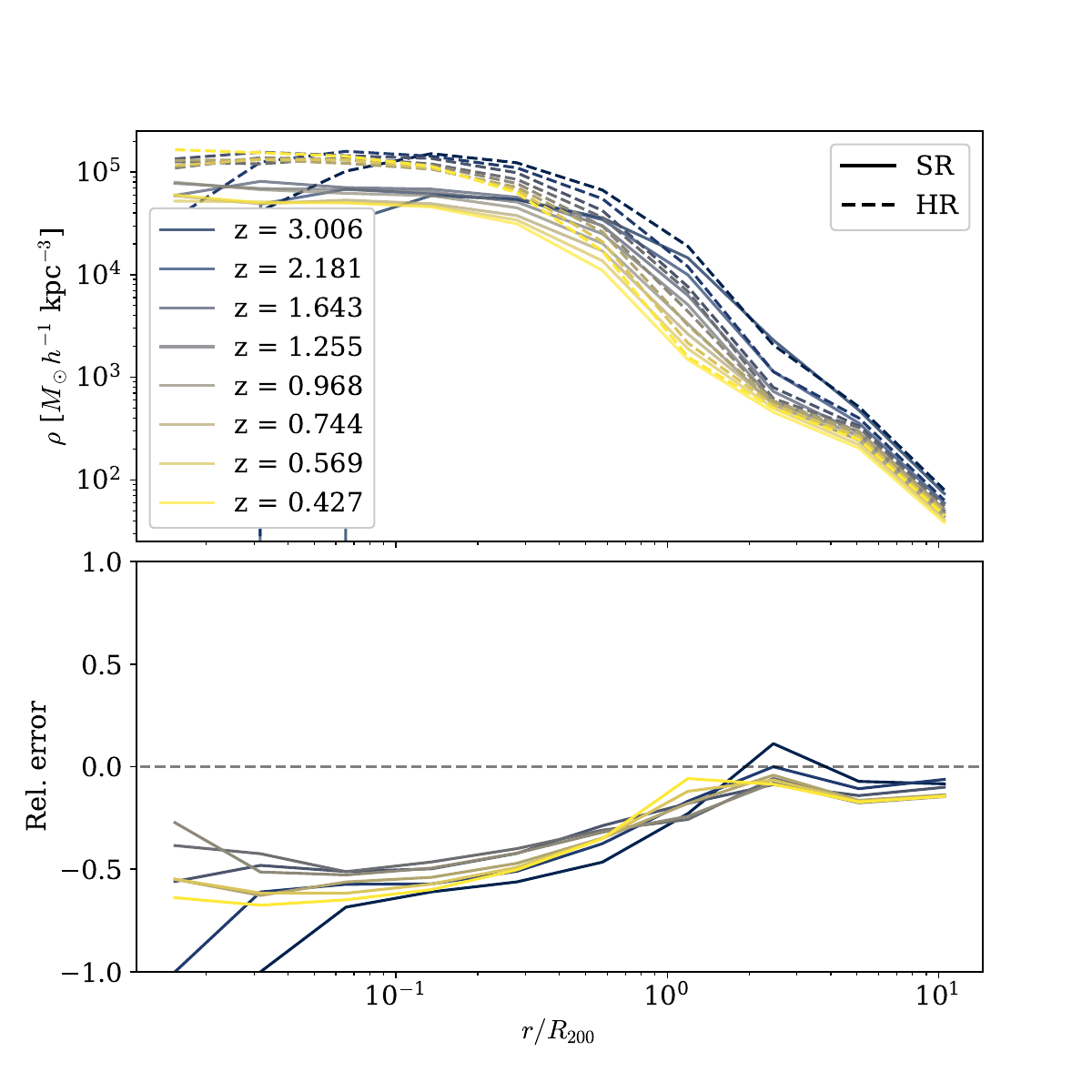}
    \caption{Halo profiles at different redshifts. The bottom panel shows the relative error between the halo profiles from the HR particles and the generated SR particles. We have only considered halos with more than $32$ particles and with mass $10^{13}M_\odot \leq M \leq 10^{14}M_\odot$.}
    \label{fig:halo_profiles}
\end{figure}

While the halo mass function characterises the abundance of halos as a function of mass, halo density profiles provide insight into their internal structure. By comparing the radial density profiles of halos identified in the super-resolved and the HR simulations, we can assess how accurately the model is able to reconstruct the internal structure of dark matter halos. This comparison is plotted in \cref{fig:halo_profiles} over a range of redshifts. We find that the model recovers the outer density profile, including the expected power law, very accurately but systematically underestimates the central density of halos, with deviations of up to $\sim 50\%$ in the innermost regions. Although the overall shape of the profiles remains broadly consistent, this suppression indicates that the model fails to fully capture the small-scale clustering required to reproduce dense halo cores. Since the inner regions of halos are resolved by relatively few particles, they are particularly sensitive to small errors in the particle positions. 

\section{Super-resolution curved universe simulation}
\label{sec:bigsim}

\begin{figure*}
    \centering \includegraphics[width=\linewidth]{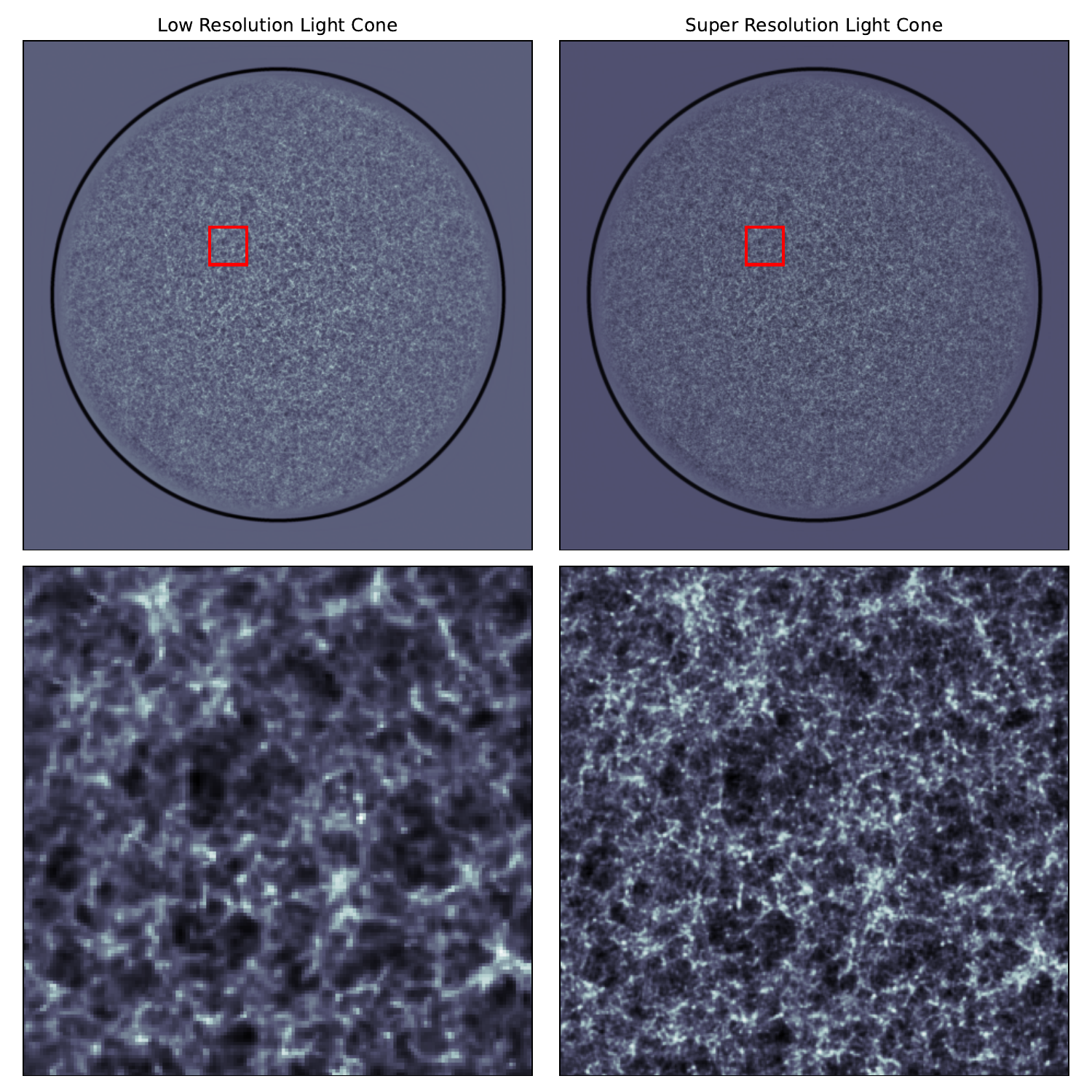}
    \caption{Slice through the density field of the low-resolution (left) and super-resolution (right) light cones. The top panels show the full $8080\,h^{-1}\,\mathrm{Mpc}$ slice where the observer is located at the centre, while the bottom panels show a zoomed-in region with side length $591.8\,h^{-1}\,\mathrm{Mpc}$. The location of the zoomed-in regions are indicated by red boxes in the top panels. The light cone is apodised and truncated at the embedding radius, beyond which the simulation code uses a flat Friedmann model to implement boundary conditions.}
    \label{fig:light-cone}
\end{figure*}

In this section, we apply our model to the particle light cone of a large-volume, LR $N$-body simulation with spatial curvature. As opposed to the separate universe boxes that we used during training, this light cone is specifically constructed to also account for the geometric effect of curvature, using the approach of \citet{Adamek:2025ylh}. This approach requires the entire survey footprint to fit into the simulation box and therefore limits the resolution one can achieve in the simulation, providing a compelling use-case scenario for our SR emulator.

Compared to the previous section where the model operated on snapshots, generating the SR output for a particle light cone is a bit more involved. One obvious difference is that the scale factor is not constant across the volume, as it depends on the distance from the observer. This is taken into account by modulating the corresponding parameter of the emulator accordingly as its spatial context window passes over the light cone. By design, the model will treat the scale factor as constant across a window of $7.8215\,h^{-1}\,\mathrm{Mpc}$ at the minimum, which is a limitation that might lead to some loss of performance.

Another complication arises due to the fact that the simulations of \citet{Adamek:2025ylh} do not use the comoving coordinates of the curved patch directly, but rather express particle positions in terms of the comoving coordinates of the flat exterior embedding space. In this context, we would like to point out that the original motivation of the separate universe approach in \citet{Li:2014sga,Wagner:2014aka} was to model the effect of a large-scale density perturbation, which is exactly how curvature is implemented in the approach of \citet{Adamek:2025ylh}. The relation between the two different comoving coordinate frames is therefore already detailed in those earlier references. In short, we may write the matter density in the curved patch as
\begin{equation} \label{eq:densityrelation}
    \tilde{\rho} = \rho \left(1+\delta_\mathrm{DC}\right)\,,
\end{equation}
where we follow the convention of \citet{Wagner:2014aka} and \citet{Adamek:2025ylh} by indicating the curved-patch quantities with a tilde. On the right-hand side, $\rho$ is the mean matter density of the flat FLRW embedding space, and $\delta_\mathrm{DC}$ is a uniform density perturbation inside the curved patch and in synchronous gauge.

Given that \cref{eq:densityrelation} holds for any time, we can write
\begin{equation} \label{eq:expansionrelation}
    \left(1 + \tilde{z}\right)^3 = \left(1 + z\right)^3 \frac{1 + \delta_\mathrm{DC}(z)}{1 + \delta_\mathrm{DC}(z=0)}\,,
\end{equation}
and therefore $\tilde{z} \approx z + {}^1\!/\!_3 (1 + z) [\delta_\mathrm{DC}(z) - \delta_\mathrm{DC}(z=0)]$. The continuity equation implies that $\tilde{H} \approx H - {}^1\!/\!_3\dot\delta_\mathrm{DC}$, and one can show that
\begin{equation}
    \tilde{\Omega}_k = -\frac{k}{\tilde{H}^2} \left(1 + \tilde{z}\right)^2 \approx -\Omega_\mathrm{m} \delta_\mathrm{DC} - \frac{2}{3 H} \dot\delta_\mathrm{DC}\,.
\end{equation}

We have to be mindful of the fact that we use Poisson gauge in our simulation. Therefore, the coordinate time is not matter synchronous and the spatial coordinates are not matter comoving. To define a displacement field that can be used by our emulator, we run a reference simulation without matter perturbations in the curved patch, providing us with a zero-point reference for the particle displacements. In other words, the particle positions of this simulation define the comoving coordinate grid of the curved patch. Furthermore, we ray trace the light cone of this simulation using the methodology of \citet{Lepori:2020ifz} on a sufficiently dense sample of particles to obtain $\tilde{z}$ as a function of coordinate distance from the observer. Together with \cref{eq:expansionrelation}, we can then infer the position-dependent conversion factor between simulation coordinates and comoving coordinates of the curved model.

The procedure for applying our SR emulator to a simulation with matter perturbations is then the following. First, we compute the coordinate distance for each particle on the light cone to its unperturbed position in the reference run. Then, we apply the position-dependent unit conversion factor to arrive at the displacement vector in comoving coordinates of the curved patch. These displacement vectors are then processed by the SR emulator, which is evaluated with a position-dependent redshift parameter. In a final step, the generated HR displacement field is converted back to simulation units to obtain the positions of the particles on the HR light cone. The results are illustrated in \cref{fig:light-cone}, where we show a side-by-side comparison of the LR simulation and the final output of our SR pipeline.

Our large-volume, LR runs have $2048^3$ particles in a box with $L_\mathrm{box} \simeq 8000\,h^{-1}\,\mathrm{Mpc}$. We use $\tilde{\Omega}_k = -0.05$, $\tilde{h} = 0.7$, $\tilde{\Omega}_\mathrm{m} = 0.32$, and for the run with matter perturbations, $\sigma_8 = 0.83$. Although they are already non-trivial by themselves, we estimate that producing these large runs requires less than $1\%$ of the computational resources (and ${}^1\!/\!_{8}$ of the disk space) that would be required to simulate the HR target directly. The reduced problem size explains only part of the savings, as most of them are due to the fact that we can run the LR simulation efficiently using a uniform mesh, without the overhead of adaptive force resolution.

\subsection{Angular power-spectrum}

The angular power spectrum $C_\ell$ provides a statistical description of the amplitude of angular fluctuations of a field defined on a sphere. Given a scalar field $f(\boldsymbol{n})$, it is defined through a spherical harmonic decomposition 
\begin{equation}
    f(\boldsymbol{n}) = \sum_{\ell m} a_{\ell m}Y_{\ell m}(\boldsymbol{n}),
\end{equation}
with $C_\ell = \langle |a_{\ell m}|^2\rangle$. While this formalism is commonly used in analyses of Cosmic Microwave Background temperature anisotropies, it can equally be applied to characterise the clustering of matter projected onto a spherical shell. In this case, the resulting angular power spectrum corresponds to a projection of the three-dimensional matter power spectrum over the radial extent of the shell. 

\begin{figure}
    \centering    \includegraphics[width=\linewidth]{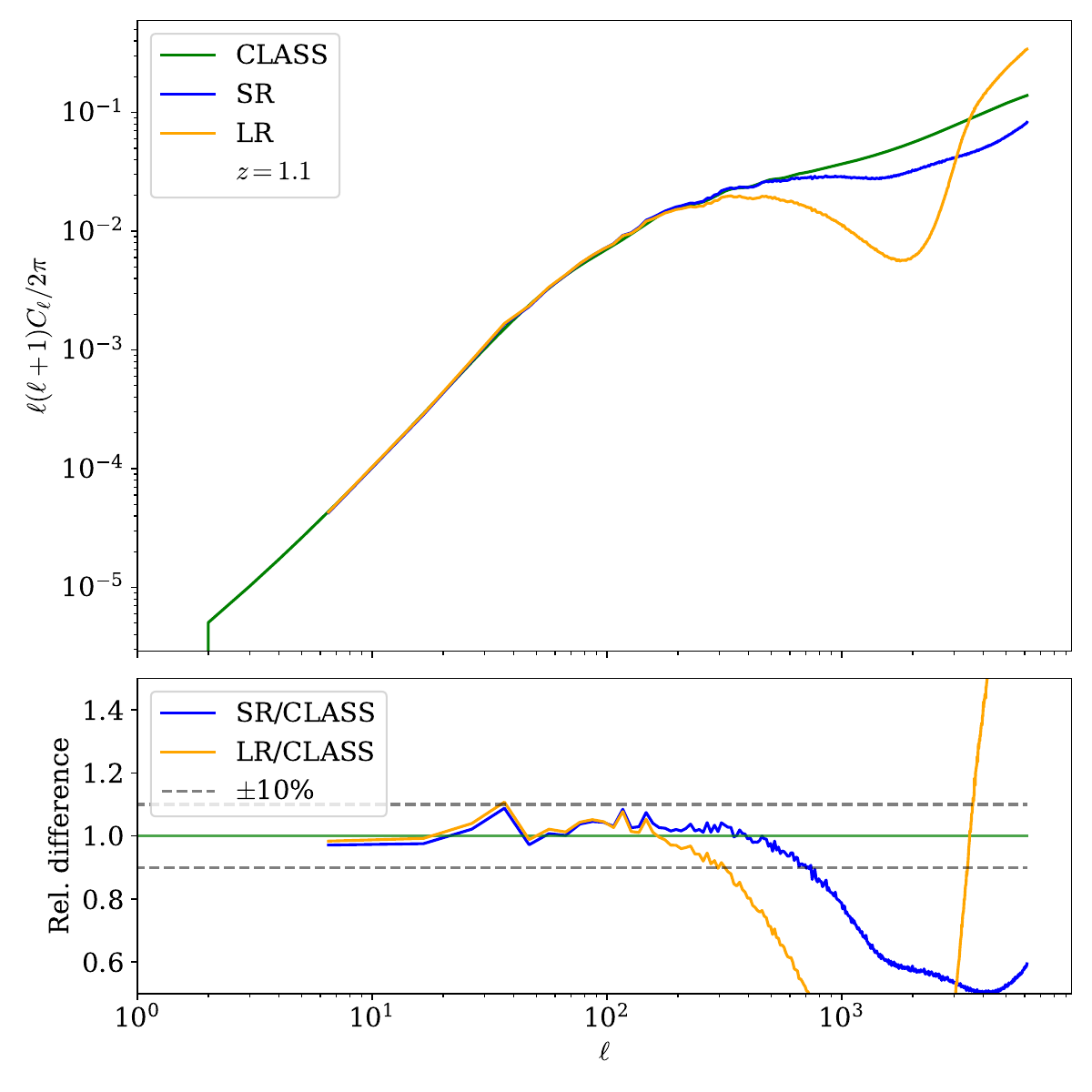}
    \caption{Angular power spectrum calculated from the SR light cone, the LR light cone and the prediction from \texttt{CLASS}, calculated at $z=1.1$. Here, we have used a redshift bin with full width $\Delta z=0.1$.
    }
    \label{fig:angular_pk}
\end{figure}

The efficacy of the upscaling can be assessed by comparing the angular power spectrum of the SR light cone to the LR one.
HR light cones are computationally expensive to generate, so the accuracy of predictions on scales not captured by the LR simulation can instead be assessed by comparison with semi-analytic estimates, such as those computed with \texttt{CLASS} \citep{Blas:2011rf,DiDio:2013bqa}. Such a comparison is presented in \cref{fig:angular_pk}. 
We compute the angular power spectrum from a spherical shell centred at $z=1.1$, with a full width of $\Delta z=0.1$. To simplify the comparison, we neglect redshift-space distortions and select particles according to their comoving distance, using the distance-redshift relation we obtained for the reference run without matter perturbations. We pixelise the density field using \texttt{HEALPix} \citep{Gorski:2004by} and estimate the binned angular power spectrum with \texttt{NaMaster}\footnote{\href{https://github.com/LSSTDESC/NaMaster}{\faicon{github}~https://github.com/LSSTDESC/NaMaster}} \citep{Alonso:2018jzx}. In the bottom panel of \cref{fig:angular_pk}, which shows the relative difference to \texttt{CLASS}, it is clear that our model is in agreement with \texttt{CLASS} within $10\%$ up to $\ell \sim 800$. This is a significant improvement over the LR result, which falls beyond $10\%$ error at $\ell \sim 300$. At higher multipoles, the accuracy of our model suffers, which is consistent with what is seen in \cref{fig:redshift-accuracy}. The predictions from \texttt{CLASS} use the nonlinear model for the matter power spectrum published by \citet{Mead:2020vgs}.

\section{Conclusion}
\label{sec:fin}

In this work, we presented a model for super-resolving displacement fields from cosmological $N$-body simulations, effectively increasing the number of particles in the system. The model is conditioned on the low-resolution displacement field, four cosmological parameters ($\Omega_k$, $\Omega_\mathrm{m}$, $\sigma_8$, and $h$), and the scale factor. We adopt a U-Net architecture, tasked with predicting the residual relative to a naively upsampled field.
Through adversarial training, the model learns to interpret injected noise as stochastic structure, enabling it to generate an ensemble of plausible high-resolution realisations. Training  the model this way allows optimising for statistical agreement rather than pixel-wise accuracy, which is well suited to the task of reconstructing unresolved small-scale structure.

The model successfully reproduces several key cosmological summary statistics, including the matter power spectrum and halo mass function, significantly increasing the effective resolution when compared to the low-resolution input. However, deviations from the ground-truth statistics arise on very small scales, where we observe a suppression of power of $10\%$-$20\%$ near the Nyquist wavenumber. This loss of small-scale structure also leads to a reduced number density of low-mass halos of up to $\sim 50\%$, as well as suppressed central densities in halo profiles. In addition, velocity reconstructions based on a finite-difference approximation successfully reproduce the velocity power spectrum, but show discrepancies on very small scales, with deviations reaching $\sim 20\%$-$40\%$ near the Nyquist wavenumber.
For $k < 1\,h\,\mathrm{Mpc}^{-1}$, the errors on density and velocity spectra remain below $\sim 10\%$ and approach $\sim 1\%$ at the scale of baryon acoustic oscillations.

We combined our post-processing pipeline with the approach by \cite{Adamek:2025ylh} to create super-resolution simulations with global spatial curvature. We created super-resolved light cones by applying the model to patches of approximately constant redshift, and tested the accuracy of this approach by comparing the measured angular power spectrum of density perturbations to the predictions of \texttt{CLASS}. 

We demonstrated that using machine learning in this way has at least two obvious benefits. First, we can create high-fidelity synthetic data at a fraction of the cost of running a full simulation at the target resolution. Second, when the training process uses different simulation frameworks for low-resolution input and high-resolution target, the model can also learn to correct for relevant differences between those two frameworks. In our example, we used a tree code with adaptive force resolution to generate the high-resolution targets, but ran a very economic fixed-grid particle-mesh method for the low-resolution input. In the future, one could imagine using this approach to incorporate other effects into the training process, with baryonic feedback among the obvious candidates.

\section*{Acknowledgements}

We thank Francisco Villaescusa-Navarro, Leander Thiele, and Yin Li for their advise on adapting the \textit{Quijote} pipeline to generate our training data.
DFM and DF thank the Research Council of Norway for their support and the resources provided by UNINETT Sigma2-the National Infrastructure for High-Performance Computing and Data Storage in Norway. 
JA acknowledges funding by the Swiss National Science Foundation and the Dr.\ Tomalla Foundation for Gravity Research. This work was supported by a grant from the Swiss National Supercomputing Centre (CSCS) under project ID sm97 on \textit{Alps}.

\section*{Data availability}

Our training data, model architecture, trained model weights, large simulation, and processing pipelines will be made public after peer review.

\bibliographystyle{aa}
\bibliography{lib}

@article{Zhang:2023lqi,
    author = "Zhang, Xiaowen and Lachance, Patrick and Ni, Yueying and Li, Yin and Croft, Rupert A. C. and Di Matteo, Tiziana and Bird, Simeon and Feng, Yu",
    title = "{AI-assisted super-resolution cosmological simulations III: time evolution}",
    eprint = "2305.12222",
    archivePrefix = "arXiv",
    primaryClass = "astro-ph.CO",
    doi = "10.1093/mnras/stad3940",
    journal = "Mon. Not. Roy. Astron. Soc.",
    volume = "528",
    number = "1",
    pages = "281--293",
    year = "2024"
}

@article{Li:2020vor,
    author = "Li, Yin and Ni, Yueying and Croft, Rupert A. C. and Di Matteo, Tiziana and Bird, Simeon and Feng, Yu",
    title = "{AI-assisted super-resolution cosmological simulations}",
    eprint = "2010.06608",
    archivePrefix = "arXiv",
    primaryClass = "astro-ph.CO",
    doi = "10.1073/pnas.2022038118",
    month = "10",
    year = "2020"
}

@article{Chisari:2018prw,
    author = "Chisari, Nora Elisa and Richardson, Mark L. A. and Devriendt, Julien and Dubois, Yohan and Schneider, Aurel and Brun, M. C., Amandine Le and Beckmann, Ricarda S. and Peirani, Sebastien and Slyz, Adrianne and Pichon, Christophe",
    title = "{The impact of baryons on the matter power spectrum from the Horizon-AGN cosmological hydrodynamical simulation}",
    eprint = "1801.08559",
    archivePrefix = "arXiv",
    primaryClass = "astro-ph.CO",
    doi = "10.1093/mnras/sty2093",
    journal = "Mon. Not. Roy. Astron. Soc.",
    volume = "480",
    number = "3",
    pages = "3962--3977",
    year = "2018"
}

@article{Huang:2018wpy,
    author = "Huang, Hung-Jin and Eifler, Tim and Mandelbaum, Rachel and Dodelson, Scott",
    title = "{Modelling baryonic physics in future weak lensing surveys}",
    eprint = "1809.01146",
    archivePrefix = "arXiv",
    primaryClass = "astro-ph.CO",
    doi = "10.1093/mnras/stz1714",
    journal = "Mon. Not. Roy. Astron. Soc.",
    volume = "488",
    number = "2",
    pages = "1652--1678",
    year = "2019"
}

@article{Euclid:2024few,
    author = "Castander, F. J. and others",
    collaboration = "Euclid",
    title = "{Euclid - V. The Flagship galaxy mock catalogue: A comprehensive simulation for the Euclid mission}",
    eprint = "2405.13495",
    archivePrefix = "arXiv",
    primaryClass = "astro-ph.CO",
    doi = "10.1051/0004-6361/202450853",
    journal = "Astron. Astrophys.",
    volume = "697",
    pages = "A5",
    year = "2025"
}

@article{Euclid:2024yrr,
    author = "Mellier, Y. and others",
    collaboration = "Euclid",
    title = "{Euclid. I. Overview of the Euclid mission}",
    eprint = "2405.13491",
    archivePrefix = "arXiv",
    primaryClass = "astro-ph.CO",
    doi = "10.1051/0004-6361/202450810",
    journal = "Astron. Astrophys.",
    volume = "697",
    pages = "A1",
    year = "2025"
}

@article{DESI:2013agm,
    author = "Levi, Michael and others",
    collaboration = "DESI",
    title = "{The DESI Experiment, a whitepaper for Snowmass 2013}",
    eprint = "1308.0847",
    archivePrefix = "arXiv",
    primaryClass = "astro-ph.CO",
    month = "8",
    year = "2013"
}

@article{Blas:2011rf,
    author = "Blas, Diego and Lesgourgues, Julien and Tram, Thomas",
    title = "{The Cosmic Linear Anisotropy Solving System (CLASS) II: Approximation schemes}",
    eprint = "1104.2933",
    archivePrefix = "arXiv",
    primaryClass = "astro-ph.CO",
    reportNumber = "CERN-PH-TH-2011-082, LAPTH-010-11",
    doi = "10.1088/1475-7516/2011/07/034",
    journal = "JCAP",
    volume = "07",
    pages = "034",
    year = "2011"
}

@article{DiDio:2013bqa,
    author = "Di Dio, Enea and Montanari, Francesco and Lesgourgues, Julien and Durrer, Ruth",
    title = "{The CLASSgal code for Relativistic Cosmological Large Scale Structure}",
    eprint = "1307.1459",
    archivePrefix = "arXiv",
    primaryClass = "astro-ph.CO",
    reportNumber = "LAPTH-036-13, CERN-PH-TH-2013-154",
    doi = "10.1088/1475-7516/2013/11/044",
    journal = "JCAP",
    volume = "11",
    pages = "044",
    year = "2013"
}

@article{Gorski:2004by,
    author = "G{\'o}rski, K. M. and Hivon, E. and Banday, A. J. and Wandelt, B. D. and Hansen, F. K. and Reinecke, M. and Bartelman, M.",
    title = "{HEALPix - A Framework for high resolution discretization, and fast analysis of data distributed on the sphere}",
    eprint = "astro-ph/0409513",
    archivePrefix = "arXiv",
    doi = "10.1086/427976",
    journal = "Astrophys. J.",
    volume = "622",
    pages = "759--771",
    year = "2005"
}

@article{Alonso:2018jzx,
    author = "Alonso, David and Sanchez, Javier and Slosar, An{\v{z}}e",
    collaboration = "LSST Dark Energy Science",
    title = "{A unified pseudo-$C_\ell$ framework}",
    eprint = "1809.09603",
    archivePrefix = "arXiv",
    primaryClass = "astro-ph.CO",
    doi = "10.1093/mnras/stz093",
    journal = "Mon. Not. Roy. Astron. Soc.",
    volume = "484",
    number = "3",
    pages = "4127--4151",
    year = "2019"
}

@article{Mead:2020vgs,
    author = {Mead, Alexander and Brieden, Samuel and Tr{\"o}ster, Tilman and Heymans, Catherine},
    title = "{hmcode-2020: improved modelling of non-linear cosmological power spectra with baryonic feedback}",
    eprint = "2009.01858",
    archivePrefix = "arXiv",
    primaryClass = "astro-ph.CO",
    doi = "10.1093/mnras/stab082",
    journal = "Mon. Not. Roy. Astron. Soc.",
    volume = "502",
    number = "1",
    pages = "1401--1422",
    year = "2021"
}

@MISC{Pylians,
    author = {{Villaescusa-Navarro}, Francisco},
    title = "{Pylians: Python libraries for the analysis of numerical simulations}",
    howpublished = {Astrophysics Source Code Library, record ascl:1811.008},
    year = 2018,
    month = nov,
    eid = {ascl:1811.008},
    pages = {ascl:1811.008},
    archivePrefix = {ascl},
    eprint = {1811.008},
    adsurl = {https://ui.adsabs.harvard.edu/abs/2018ascl.soft11008V}
}

@article{Villaescusa-Navarro:2019bje,
    author = "Villaescusa-Navarro, Francisco and others",
    title = "{The Quijote simulations}",
    eprint = "1909.05273",
    archivePrefix = "arXiv",
    primaryClass = "astro-ph.CO",
    doi = "10.3847/1538-4365/ab9d82",
    journal = "Astrophys. J. Suppl.",
    volume = "250",
    number = "1",
    pages = "2",
    year = "2020"
}

@article{LSST:2008ijt,
    author = "Ivezi{\'c}, {\v{Z}}eljko and others",
    collaboration = "LSST",
    title = "{LSST: from Science Drivers to Reference Design and Anticipated Data Products}",
    eprint = "0805.2366",
    archivePrefix = "arXiv",
    primaryClass = "astro-ph",
    reportNumber = "SLAC-PUB-16076",
    doi = "10.3847/1538-4357/ab042c",
    journal = "Astrophys. J.",
    volume = "873",
    number = "2",
    pages = "111",
    year = "2019"
}

@article{Newburgh:2016mwi,
    author = "Newburgh, L. B. and others",
    editor = "Hall, Helen J. and Gilmozzi, Roberto and Marshall, Heather K.",
    title = "{HIRAX: A Probe of Dark Energy and Radio Transients}",
    eprint = "1607.02059",
    archivePrefix = "arXiv",
    primaryClass = "astro-ph.IM",
    doi = "10.1117/12.2234286",
    journal = "Proc. SPIE Int. Soc. Opt. Eng.",
    volume = "9906",
    pages = "99065X",
    year = "2016"
}

@article{SKA:2018ckk,
    author = "Bacon, David J. and others",
    collaboration = "SKA",
    title = "{Cosmology with Phase 1 of the Square Kilometre Array: Red Book 2018: Technical specifications and performance forecasts}",
    eprint = "1811.02743",
    archivePrefix = "arXiv",
    primaryClass = "astro-ph.CO",
    doi = "10.1017/pasa.2019.51",
    journal = "Publ. Astron. Soc. Austral.",
    volume = "37",
    pages = "e007",
    year = "2020"
}

@article{Planck:2018vyg,
    author = "Aghanim, N. and others",
    collaboration = "Planck",
    title = "{Planck 2018 results. VI. Cosmological parameters}",
    eprint = "1807.06209",
    archivePrefix = "arXiv",
    primaryClass = "astro-ph.CO",
    doi = "10.1051/0004-6361/201833910",
    journal = "Astron. Astrophys.",
    volume = "641",
    pages = "A6",
    year = "2020",
    note = "[Erratum: Astron. Astrophys. 652, C4 (2021)]"
}

@article{DESI:2025zgx,
    author = "Abdul Karim, M. and others",
    collaboration = "DESI",
    title = "{DESI DR2 results. II. Measurements of baryon acoustic oscillations and cosmological constraints}",
    eprint = "2503.14738",
    archivePrefix = "arXiv",
    primaryClass = "astro-ph.CO",
    reportNumber = "FERMILAB-PUB-25-0169-PPD",
    doi = "10.1103/tr6y-kpc6",
    journal = "Phys. Rev. D",
    volume = "112",
    number = "8",
    pages = "083515",
    year = "2025"
}

@article{DiDio:2016ykq,
    author = "Di Dio, Enea and Montanari, Francesco and Raccanelli, Alvise and Durrer, Ruth and Kamionkowski, Marc and Lesgourgues, Julien",
    title = "{Curvature constraints from Large Scale Structure}",
    eprint = "1603.09073",
    archivePrefix = "arXiv",
    primaryClass = "astro-ph.CO",
    reportNumber = "HIP-2016-14-TH",
    doi = "10.1088/1475-7516/2016/06/013",
    journal = "JCAP",
    volume = "06",
    pages = "013",
    year = "2016"
}

@article{Li:2014sga,
    author = "Li, Yin and Hu, Wayne and Takada, Masahiro",
    title = "{Super-Sample Covariance in Simulations}",
    eprint = "1401.0385",
    archivePrefix = "arXiv",
    primaryClass = "astro-ph.CO",
    doi = "10.1103/PhysRevD.89.083519",
    journal = "Phys. Rev. D",
    volume = "89",
    number = "8",
    pages = "083519",
    year = "2014"
}

@article{Wagner:2014aka,
    author = "Wagner, Christian and Schmidt, Fabian and Chiang, Chi-Ting and Komatsu, Eiichiro",
    title = "{Separate Universe Simulations}",
    eprint = "1409.6294",
    archivePrefix = "arXiv",
    primaryClass = "astro-ph.CO",
    doi = "10.1093/mnrasl/slu187",
    journal = "Mon. Not. Roy. Astron. Soc.",
    volume = "448",
    number = "1",
    pages = "L11--L15",
    year = "2015"
}

@article{Adamek:2025ylh,
    author = "Adamek, Julian and Boschetti, Renan",
    title = "{Incorporating curved geometry in cosmological simulations}",
    eprint = "2508.20606",
    archivePrefix = "arXiv",
    primaryClass = "gr-qc",
    doi = "10.1088/1475-7516/2026/03/013",
    journal = "JCAP",
    volume = "03",
    pages = "013",
    year = "2026"
}

@article{Chen:2025mlf,
    author = "Chen, Shi-Fan and Zaldarriaga, Matias",
    title = "{It's all Ok: curvature in light of BAO from DESI DR2}",
    eprint = "2505.00659",
    archivePrefix = "arXiv",
    primaryClass = "astro-ph.CO",
    doi = "10.1088/1475-7516/2025/08/014",
    journal = "JCAP",
    volume = "08",
    pages = "014",
    year = "2025"
}

@article{Lepori:2020ifz,
    author = "Lepori, Francesca and Adamek, Julian and Durrer, Ruth and Clarkson, Chris and Coates, Louis",
    title = "{Weak-lensing observables in relativistic N-body simulations}",
    eprint = "2002.04024",
    archivePrefix = "arXiv",
    primaryClass = "astro-ph.CO",
    doi = "10.1093/mnras/staa2024",
    journal = "Mon. Not. Roy. Astron. Soc.",
    volume = "497",
    number = "2",
    pages = "2078--2095",
    year = "2020"
}

@article{Lewis:1999bs,
    author = "Lewis, Antony and Challinor, Anthony and Lasenby, Anthony",
    title = "{Efficient computation of CMB anisotropies in closed FRW models}",
    eprint = "astro-ph/9911177",
    archivePrefix = "arXiv",
    doi = "10.1086/309179",
    journal = "Astrophys. J.",
    volume = "538",
    pages = "473--476",
    year = "2000"
}

@article{Adamek:2015eda,
    author = "Adamek, Julian and Daverio, David and Durrer, Ruth and Kunz, Martin",
    title = "{General relativity and cosmic structure formation}",
    eprint = "1509.01699",
    archivePrefix = "arXiv",
    primaryClass = "astro-ph.CO",
    doi = "10.1038/nphys3673",
    journal = "Nature Phys.",
    volume = "12",
    pages = "346--349",
    year = "2016"
}

@article{Adamek:2016zes,
    author = "Adamek, Julian and Daverio, David and Durrer, Ruth and Kunz, Martin",
    title = "{gevolution: a cosmological N-body code based on General Relativity}",
    eprint = "1604.06065",
    archivePrefix = "arXiv",
    primaryClass = "astro-ph.CO",
    doi = "10.1088/1475-7516/2016/07/053",
    journal = "JCAP",
    volume = "07",
    pages = "053",
    year = "2016"
}

@article{Springel:2020plp,
    author = {Springel, Volker and Pakmor, R{\"u}diger and Zier, Oliver and Reinecke, Martin},
    title = "{Simulating cosmic structure formation with the gadget-4 code}",
    eprint = "2010.03567",
    archivePrefix = "arXiv",
    primaryClass = "astro-ph.IM",
    doi = "10.1093/mnras/stab1855",
    journal = "Mon. Not. Roy. Astron. Soc.",
    volume = "506",
    number = "2",
    pages = "2871--2949",
    year = "2021"
}

@article{Rouhiainen:2023ewv,
    author = {Rouhiainen, Adam and M{\"u}nchmeyer, Moritz and Shiu, Gary and Gira, Michael and Lee, Kangwook},
    title = "{Superresolution emulation of large cosmological fields with a 3D conditional diffusion model}",
    eprint = "2311.05217",
    archivePrefix = "arXiv",
    primaryClass = "astro-ph.CO",
    doi = "10.1103/PhysRevD.109.123536",
    journal = "Phys. Rev. D",
    volume = "109",
    number = "12",
    pages = "123536",
    year = "2024"
}

@article{Ramanah:2020vyl,
    author = "Ramanah, Doogesh Kodi and Charnock, Tom and Villaescusa-Navarro, Francisco and Wandelt, Benjamin D.",
    title = "{Super-resolution emulator of cosmological simulations using deep physical models}",
    eprint = "2001.05519",
    archivePrefix = "arXiv",
    primaryClass = "astro-ph.CO",
    doi = "10.1093/mnras/staa1428",
    journal = "Mon. Not. Roy. Astron. Soc.",
    volume = "495",
    pages = "4227",
    year = "2020"
}

@article{Sipp:2022tdp,
    author = "Sipp, Meris and LaChance, Patrick and Croft, Rupert and Ni, Yueying and Di Matteo, Tiziana",
    title = "{Towards super-resolution simulations of the fuzzy dark matter cosmological model}",
    eprint = "2210.12907",
    archivePrefix = "arXiv",
    primaryClass = "astro-ph.CO",
    doi = "10.1093/mnras/stad2341",
    journal = "Mon. Not. Roy. Astron. Soc.",
    volume = "525",
    number = "2",
    pages = "1807--1813",
    year = "2023"
}

@article{Goodfellow:2014upx,
    author = "Goodfellow, Ian J. and Pouget-Abadie, Jean and Mirza, Mehdi and Xu, Bing and Warde-Farley, David and Ozair, Sherjil and Courville, Aaron and Bengio, Yoshua",
    title = "{Generative Adversarial Networks}",
    eprint = "1406.2661",
    archivePrefix = "arXiv",
    primaryClass = "stat.ML",
    month = "6",
    year = "2014"
}

@article{Arjovsky:2017fjr,
    author = "Arjovsky, Martin and Chintala, Soumith and Bottou, L{\'e}on",
    title = "{Wasserstein GAN}",
    eprint = "1701.07875",
    archivePrefix = "arXiv",
    primaryClass = "stat.ML",
    month = "1",
    year = "2017"
}

@article{Gulrajani:2017kpy,
    author = "Gulrajani, Ishaan and Ahmed, Faruk and Arjovsky, Martin and Dumoulin, Vincent and Courville, Aaron",
    title = "{Improved Training of Wasserstein GANs}",
    eprint = "1704.00028",
    archivePrefix = "arXiv",
    primaryClass = "cs.LG",
    month = "3",
    year = "2017"
}

@article{Ni:2021mzk,
    author = "Ni, Yueying and Li, Yin and Lachance, Patrick and Croft, Rupert A. C. and Di Matteo, Tiziana and Bird, Simeon and Feng, Yu",
    title = "{AI-assisted superresolution cosmological simulations {\textendash} II. Halo substructures, velocities, and higher order statistics}",
    eprint = "2105.01016",
    archivePrefix = "arXiv",
    primaryClass = "astro-ph.CO",
    doi = "10.1093/mnras/stab2113",
    journal = "Mon. Not. Roy. Astron. Soc.",
    volume = "507",
    number = "1",
    pages = "1021--1033",
    year = "2021"
}

@article{Efstathiou:1985re,
    author = "Efstathiou, G. and Davis, M. and Frenk, C. S. and White, Simon D. M.",
    title = "{Numerical Techniques for Large Cosmological N-Body Simulations}",
    reportNumber = "NSF-ITP-84-108",
    doi = "10.1086/191003",
    journal = "Astrophys. J. Suppl.",
    volume = "57",
    pages = "241--260",
    year = "1985"
}

@software{yt.astro.analysis,
  author       = {Britton Smith and
                  Matthew Turk and
                  John ZuHone and
                  Clément Robert and
                  Stephen Skory and
                  Cameron Hummels and
                  Andrew Myers and
                  Kacper Kowalik and
                  Corentin Cadiou and
                  eganhila and
                  Sam Skillman and
                  Michael S. Warren and
                  John Wise and
                  gsiisg and
                  madcpf and
                  Sam Leitner and
                  Anthony Scopatz and
                  Miguel de Val-Borro and
                  Casey W. Stark and
                  Ho, Meng-Yuan and
                  Ben Keller and
                  Bili Dong and
                  Mark Richardson and
                  Matthew Scott Krafczyk and
                  Nathan Goldbaum and
                  Sriram Sankar and
                  stonnes},
  title        = {yt-project/yt\_astro\_analysis: yt\_astro\_analysis
                   1.1.4 Release
                  },
  month        = sep,
  year         = 2025,
  publisher    = {Zenodo},
  version      = {yt\_astro\_analysis-1.1.4},
  doi          = {10.5281/zenodo.17106789},
  url          = {https://doi.org/10.5281/zenodo.17106789},
}

@ARTICLE{yt,
   author = {{Turk}, M.~J. and {Smith}, B.~D. and {Oishi}, J.~S. and {Skory}, S. and
{Skillman}, S.~W. and {Abel}, T. and {Norman}, M.~L.},
    title = "{yt: A Multi-code Analysis Toolkit for Astrophysical Simulation Data}",
  journal = {The Astrophysical Journal Supplement Series},
archivePrefix = "arXiv",
   eprint = {1011.3514},
 primaryClass = "astro-ph.IM",
     year = 2011,
    month = jan,
   volume = 192,
      eid = {9},
    pages = {9},
      doi = {10.1088/0067-0049/192/1/9},
   adsurl = {http://adsabs.harvard.edu/abs/2011ApJS..192....9T}
}

\end{document}